\documentclass[useAMS]{mn2e}
\usepackage{times}
\usepackage{psfig}

\def\gsim{ \lower .75ex \hbox{$\sim$} \llap{\raise .27ex \hbox{$>$}} } 
\def\lsim{ \lower .75ex\hbox{$\sim$} \llap{\raise .27ex \hbox{$<$}} }

\def\ama{$E_{\rm peak}-E_{\rm iso}$}
\def\amaf{$E^{\rm obs}_{\rm peak}$--Fluence}
\def\sw{{\it Swift}}
\def\sax{{\it Beppo}SAX}
\def\he{{\it Hete-II}}
\def\ko{{\it Konus-Wind}}
\def\su{{\it Suzaku}}
\def\ba{BATSE}
\def\ep{$E_{\rm peak}$}
\def\epo{$E^{\rm obs}_{\rm peak}$}
\def\epof{$E^{\rm obs}_{\rm peak}$--Fluence}
\def\eiso{$E_{\rm iso}$}

\title[The \ama\ and the \amaf\ planes]
{The \ama\ plane of long Gamma Ray Bursts and selection effects}

\author[G. Ghirlanda, L. Nava, G. Ghisellini, C. Firmani, J.I. Cabrera]
{G. Ghirlanda$^1$\thanks{E--mail: giancarlo.ghirlanda@brera.inaf.it}, 
L. Nava$^{1,2}$, G. Ghisellini$^1$, 
C. Firmani$^{1,3}$, J. I. Cabrera$^{3}$\\
$^1$ Osservatorio Astronomico di Brera, via Bianchi 46, Merate  Italy  \\ 
$^2$ Univ. dell'Insubria, Como, Italy \\
$^3$ Instituto de Astronom\'{\i}a, U.N.A.M., A.P. 70-264, 04510, 
M\'exico, D.F., M\'exico }

\begin{document}

\pagerange{\pageref{firstpage}--\pageref{lastpage}} \pubyear{2002}
\maketitle

\begin{abstract}
  We study the distribution of long Gamma Ray Bursts in the \ama\ and
  in the \amaf\ planes through an updated sample of 76 bursts, with
  measured redshift and spectral parameters, detected up to September
  2007.  We confirm the existence of a strong rest frame correlation
  $E_{\rm peak} \propto E_{\rm iso}^{0.54\pm 0.01}$.  Contrary to
  previous studies, no sign of evolution with redshift of the
  \ama\ correlation (either its slope and normalisation) is found.
  The 76 bursts define a strong \epof\ correlation in the observer
  frame ($E^{\rm obs}_{\rm peak}\propto F_{\rm bol}^{0.32\pm 0.05}$)
  with redshifts evenly distributed along this correlation.  We study
  possible instrumental selection effects in the observer frame
  \epof\ plane.  In particular, we concentrate on the minimum peak
  flux necessary to trigger a given GRB detector (trigger threshold)
  and the minimum fluence a burst must have to determine the value of
  $E^{\rm obs}_{\rm peak}$ (spectral analysis threshold).  We find
  that the latter dominates in the \epof\ plane over the former. Our
  analysis shows, however, that these instrumental selection effects
  do not dominate for bursts detected before the launch of the
  \sw\ satellite, while the spectral analysis threshold is the
  dominant truncation effect of the \sw\ GRB sample (27 out of 76
  events). This suggests that the \epof\ correlation defined by the
  pre--\sw\ sample could be affected by other, still not understood,
  selection effects. Besides we caution about the conclusions on the
  existence of the \epof\ correlation based on our \sw\ sample alone
\end{abstract} 

\section{Introduction}

One of the properties of long Gamma Ray Bursts (GRBs) that remains mysterious,
but potentially fundamental for understanding their physics, is the observed
correlation between the bolometric energy \eiso\ emitted during their prompt
emission and the the peak of the spectrum \ep\ in a $\nu F_\nu$ plot.  In the
observer frame a correlation between the total fluence and \ep\ was found by
Lloyd, Petrosian \& Mallozzi (2000, LPM00 hereafter) with a sample of \ba\ 
bursts without measured redshifts. In their pioneering work, LPM00 predicted
the existence of an intrinsic $E_{\rm peak}\propto E_{\rm iso}^{a}$
correlation with $a\in[0.47,0.62]$. Later, Amati et al. (2002) indeed found
such a correlation, based on a sample of 12 GRBs observed by the \sax\ 
satellite and with spectroscopically measured redshifts.  The following
updates (e.g. Lamb, Donaghy \& Graziani 2005; Amati 2006; Ghirlanda et al.
2007) confirmed this correlation and showed that the exponent depends somewhat
on the fitting method and on the sample under consideration, but is of the
order of $a\sim 0.5$.


The current debate about the \ama\ correlation concerns (a) the very existence
of the correlation and the presence of outliers; (b) the evolution with
redshift of the slope and normalisation of the correlation and (c) the
presence of selection effects on this correlation.

The real existence of the Amati correlation has been questioned by Nakar \&
Piran (2005) and Band \& Preece (2005) who considered different samples of
GRBs, detected by BATSE, of known \ep\ but without redshift determination.  By
considering all possible redshifts, they claimed that a large fraction of GRBs
were in any case outliers to the {\it original} (Amati et al. 2002) Amati
relation.  They also claimed that the \ama\ correlation was a boundary of a
larger dispersion of points in the rest frame \ama\ plane. Ghirlanda et al.
(2005), using a large sample of 442 GRBs with pseudo--redshifts determined by
the lag--luminosity relation (Norris, Marani \& Bonnell 2000; Norris 2002;
Band \& Preece 2005), found that \ep\ and \eiso\ strongly correlate.  The
correlation has a normalisation and scatter slightly larger than in the
original Amati et al. (2002) paper.  Although these pseudo--redshifts are
based on the still uncertain lag--luminosity correlation, they could be used
to show that there is no outlier of the \ama\ correlation within the used
sample of 442 GRB.  Similar conclusions were reached by Bosnjak et al.
(2007), using a different method. Consider that in all cases two GRBs (GRB
980425 and GRB 031203) were not included in the fits, being clear outliers of
the correlation and also anomalous in many ways, even if there have been
attempts to make them mainstream by considering them off--axis events
(Ramirez--Ruiz et al. 2005), or sources whose prompt flux is dimmed by some
scattering material, or spectral evolution (Ghisellini et al.  2006).

\begin{figure*}
\centerline{\psfig{figure=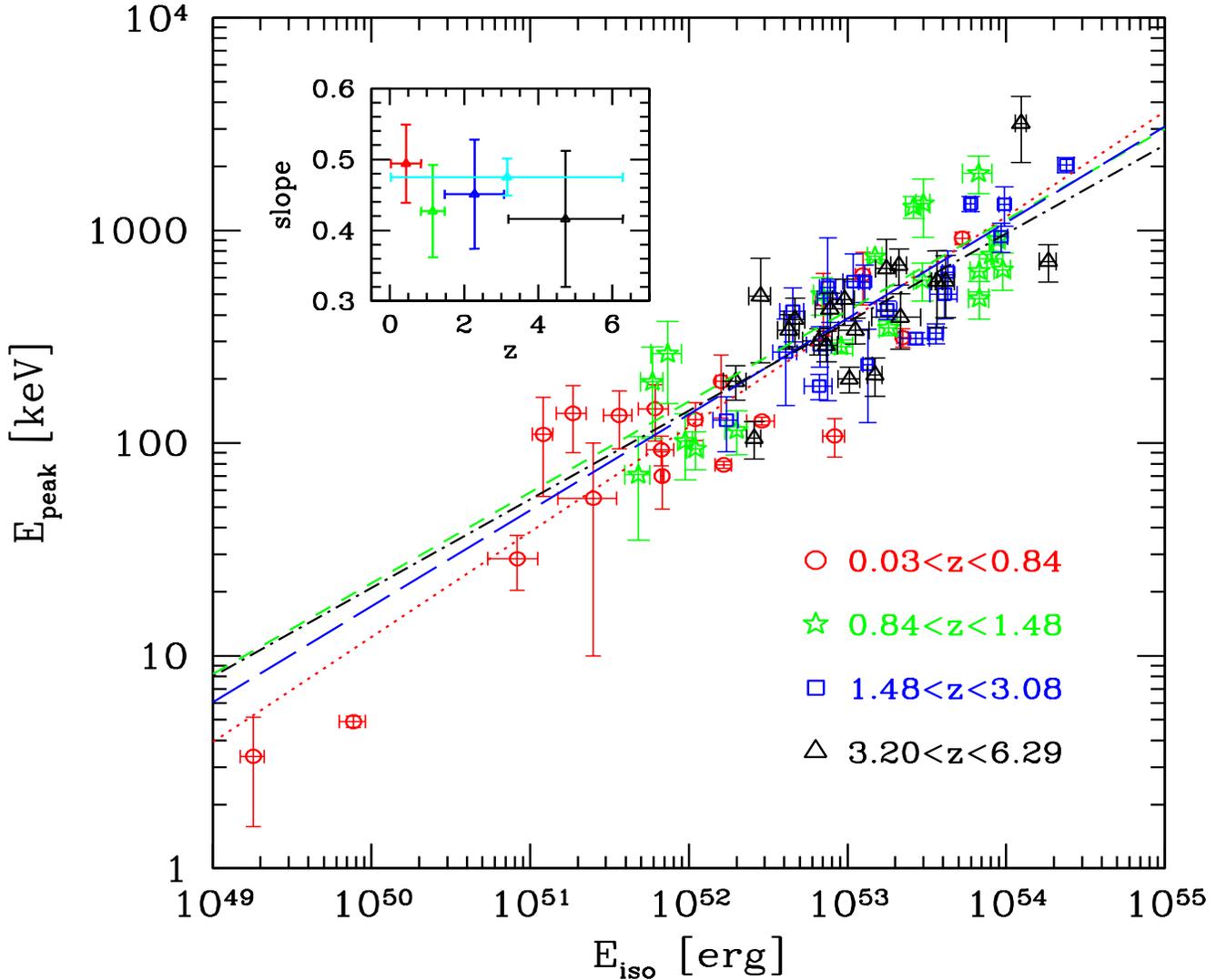,width=19cm,height=17cm}}
\vskip -1 true cm
\caption{ The 76 GRBs  with known redshifts and well measured spectral
  properties (updated to September 2007) in the rest frame plane.  Bursts are
  divided in four redshift bins (as labelled).  The lines are the best fit
  obtained with the least squares method (dotted, short--dashed, long--dashed,
  dot--dashed from small to high redshift values, respectively).  The insert
  represents the slope as a function of the 4 redshift bin.  The slope of the
  correlation defined with the entire sample of 76 bursts is also shown (cyan
  symbol in the insert). }
\label{amati}
\end{figure*}

Li (2007, hereafter L07), considering a sample of 48 GRBs (from Amati 2006;
2007) investigated if the correlation evolves with redshift, finding that it
does, becoming steeper (i.e. larger $a$) at higher redshifts.

Very recently, Butler et al. (2007, hereafter B07) claimed that the
correlation exists, but it is probably the result of a selection effect,
similar to the correlations often found when considering flux limited samples
and calculating the correlation between the luminosities in different bands.
They suggest that by multiplying the fluence and the observed peak energy by
strong function of redshift can induce a correlation in the rest frame.

In this work we study these issues by updating the sample of bursts with known
redshifts (\S 2). We study the evolution of the \ama\ correlation with
redshift in \S 3. In \S 4 we show the existence of a strong \epof\ 
correlation.  The GRB sample considered is heterogeneous in terms of the
instruments that detected the bursts. This requires a deteailed analysis of
the possible selection effects in order to understand if the \ama\ correlation
is an intrinsic correlation or if it is due to any selection effect. We study
the instrumental selection effects in the \epof\ observer frame and discuss
their impact on the \ama\ correlation (\S 4).  The relation between the \epof\ 
and the \ama\ relation is briefly discussed in \S 5, and in \S 6 we draw our
conclusions.

In this work we focus on the sample of bursts with measured redshifts and well
defined spectral properties in order to study the issue of the redshift
evolution (L07) and the selection effects on the \ama\ correlation (see also
B07). We will study the observer frame \epof\ plane with larger samples of
bursts of unknown $z$ in a forthcoming paper (Nava et al. in preparation).

We use $H_0=70$ km s$^{-1}$ Mpc$^{-1}$, $\Omega_{\rm M}=0.3$ and
$\Omega_\Lambda=0.7$.

\begin{figure}
\hskip -0.7 true cm
{\psfig{figure=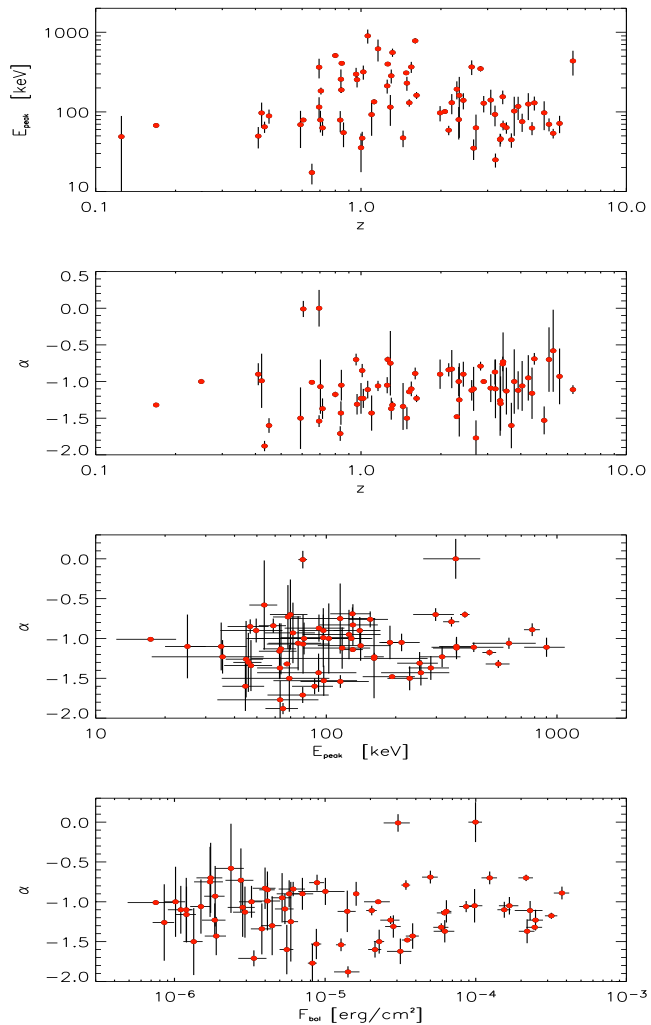,width=9cm,height=14cm}} 
\caption{Correlation of the spectral parameters  (\epo and $\alpha$) with
  redshift (top panels) and between the spectral parameters (bottom left)
  and the fluence (bottom right) for the sample of bursts with known redshift
  reported in Tab. \ref{tab1}.}
\label{correle}
\end{figure}
\begin{figure}
\hskip -0.7 true cm
{\psfig{figure=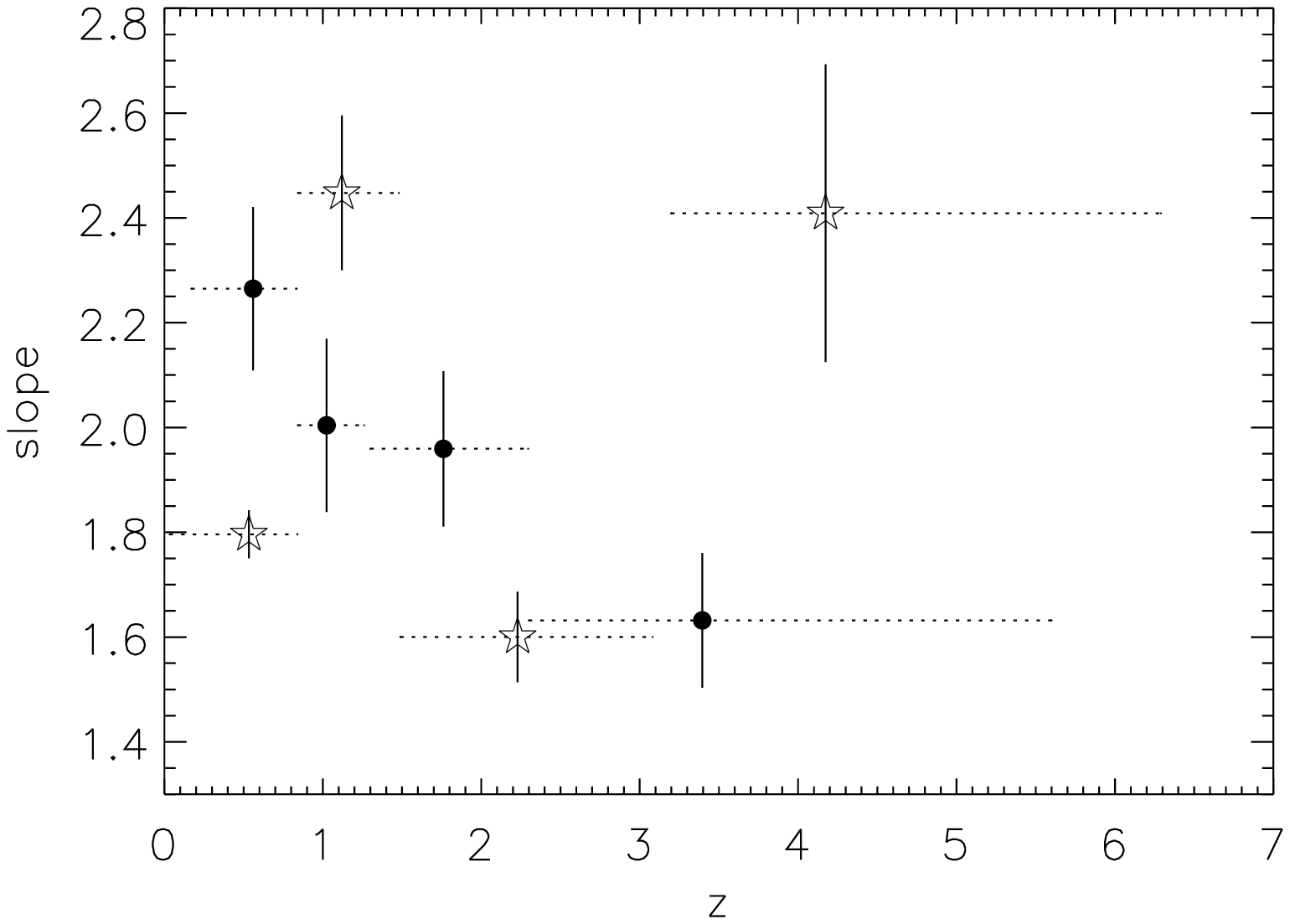,width=9cm,height=8cm}} 
\caption{Evolution of the slope of the \ama\ correlation (defined as $E_{\rm iso}\propto E_{\rm peak}^{1/a}$ with redshift. The circles are the results
  obtained with the 48 GRBs of the L07 sample and the open stars are the
  results obtained with the 76 GRBs of our sample. The fit were performed by
  weighting for the erros on both \ep\ and \eiso\ for consistency with the
  method adopted by L07.}
\label{li}
\end{figure}

\section{The \ama\ plane}

The information required to put a GRB on the \ama\ plane are (i) the redshift,
(ii) the spectral parameters and (iii) the fluence. These are used to compute
the bolometric isotropic energy $E_{\rm iso}$ (e.g. see Ghirlanda, Ghisellini
\& Lazzati 2004) and the rest frame peak energy $E_{\rm peak}$ of the $\nu
F_{\nu}$ spectrum. Note that the fluence and peak energy are of the time
integrated spectrum, i.e. integrated over the total duration of the burst.

We have collected all bursts with spectroscopically measured redshift
and with known spectral properties.  35 GRBs were detected by
instruments on-board different satellites ({\it CGRO}, {\it Beppo}SAX,
{\it Hete--II}) before the \sw\ satellite was launched in Nov. 2004
(Gehrels et al. 2004) and 41 events were detected by different
satellites ({\it Konus--Wind, Swift, RHESSI, Suzaku, {\it Hete--II} })
since the end of 2004 in the so called ``\sw\ era''. Among the latter
in 27 cases the spectral parameters (peak energy and fluence) were
derived from the analysis of the \sw--BAT data. For 19 out of 27
bursts the spectral parameters were taken from the compilation of
Cabrera et al. 2007 (hereafter C07) and for the other 8 bursts the
spectral parameters are from the literature\footnote{Note that almost
  all the 41 bursts detected since 2005 were detected also by
  \sw. However, only in 27 cases the \sw--BAT spectrum could constrain
  the peak energy (C07). In all the other cases only the detection by the
  other satellites ({\it Konus--Wind, RHESSI, Suzaku}) allowed to
  constrain \epo.}. We refer to this sample of 27 bursts as the
``\sw\ burst sample'' and define the sample of all the other 49 bursts
the non--\sw\ sample.

In Tab. \ref{tab1} we report the redshifts, spectral properties and isotropic
energy of the sample of the 76 GRBs.  This is the most updated sample of
bursts with published redshift, \ep\ and fluence that can be put in the \ama\ 
plane up to September 2007.  For many bursts the time integrated spectrum was
fitted with the Band model (Band et al. 1993) or with a cutoff power--law.
There are indications (e.g.  Kaneko et al. 2006, hereafter K06) that if a
sufficiently broad energy range is covered by the available spectrum (e.g. up
to few MeV), the {\it time integrated} spectrum is preferentially represented
by the Band model.  In order to uniformly estimate $E_{\rm iso}$, following
Ghirlanda et al.  (2007), for those bursts with a spectrum fitted by a cutoff
power--law we computed the logarithmic average of \eiso\ derived with this
model and the value derived with the Band function by fixing the high energy
spectral index to --2.3 (i.e. the typical value reported in e.g. K06).  For
the GRBs taken from C07, {\ep\ was transformed into linear (and its error
  symmetrized), in order to have the same format of several other tables
  already published in the literature.  Note however that for the points
  discussed below, our fits do not weight for the errors on the variables.
  All the burst before 2005 taken from Amati (2006) are corrected for the
  different assumption on $H_0$ (Amati 2006 used $H_0=65$ km s$^{-1}$
  Mpc$^{-1}$).
  
  In Fig. \ref{amati} we show the correlation defined by the 76 bursts listed
  in Tab. \ref{tab1}.  Statistical analysis gives a Kendall's tau correlation
  coefficient $\tau=0.68$ (18$\sigma$ significance).  The correlation can be
  modelled with a power law.  Actually, different fitting procedures have been
  adopted in the literature: (a) simple least square fit which minimizes the
  difference between the the data points and the model (along the ordinate
  direction) without wighting for the errors of the data points or (b) a fit
  that weights for the errors on both variables (see e.g. Press et al. 1986).
  The least square fit gives $\log E_{\rm peak}= (-22.59\pm1.39)+ \log E_{\rm
    iso}^{0.47\pm0.03}$ with a $\chi^2=3.49$. The fit obtained by weighting
  for the errors is $\log E_{\rm peak}= (-26.66\pm0.46)+ \log E_{\rm
    iso}^{0.55\pm0.01}$ with a $\chi^2=545$ for 74 degrees of freedom.  In the
  latter case, the reduced $\chi^2$ is extremely large and this is due to the
  sample dispersion (see below) which is much larger than the statistical
  errors associated with the variables.  Such a large $\chi^2$ also leads to
  underestimate the errors of the parameters of the fit.  It has been proposed
  (Reichart et al.  2001, 2005, but see Guidorzi et al. 2005, see also
  D'agostini et al. 2006) that the fit should account for an additional
  parameter which is the sample variance.  Amati (2006) performed this fit on
  his GRB sample and found a quite large value of the sample variance
  $\sigma=0.15$.  Instead, we proposed (Ghirlanda et al. 2004, see also Liang
  \& Zhang 2005) that the large dispersion of the data points in the \ama\ 
  plane is due to a third observable, i.e. the jet break time.  Indeed, by
  correcting the isotropic energy for the collimation angle that can be
  derived from the jet break time of the optical light curve, the dispersion
  of the \ama\ correlation is greatly reduced (Ghirlanda et al. 2007, Nava et
  al. 2007).  In this work we will adopt the least square fitting method
  (Press et al. 1986 p.499) without weighting for the errors on the variables
  because the sample dispersion in the \ama\ plane defined by the 76 GRBs
  ($\sigma$=0.22 - consistent with the values found with smaller GRB samples,
  e.g. Ghirlanda et al. 2004, 2005; Amati et al. 2006, C07) is clearly larger
  than the typical statistical errors associated with either \ep\ and \eiso
  ($\langle \sigma_{logE_{\rm p}} \rangle =0.1$ and $\langle \sigma_{logE_{\rm
      iso}} \rangle =0.06$).
  
  With the most updated sample of 76 GRBs reported in Tab. \ref{tab1} we can
  also verify that no correlation exists between the spectral parameters or
  with the redshift (see Fig. \ref{correle}). In particular, there is no
  correlation between $\alpha$ and $z$ nor between $\alpha$ and \epo (second
  and third panel in Fig.~\ref{correle}). A different, strong, correlation
  between $\alpha_{\rm pl}$ of the fit with a single power law and the peak
  energy of the fit with a cutoff--power law has been reported recently
  (Sakamoto et al. 2006; Butler et al. 2007) from the analysis of the \sw-BAT
  spectra. Note that in this correlation the two spectral parameters
  $\alpha_{\rm pl}$ and \epo\ belong to two different models, i.e. a single
  power law and a cutoff--power law model, respectively.  However, it has been
  shown in C07 that this correlation is spurious (Fig.  3 and Sec. 3.2 of that
  paper).  Although a correlation indeed appears between the single power law
  spectral index $\alpha_{pl}$ and \epo\ (of the cutoff--power law model --
  see bottom panel of Fig. 3 in C07), it has no physical meaning: it is the
  result of the attempt of the single power law model fit to account for the
  spectral data above the peak which have a smaller $\nu F_{\nu}$ flux.
  Spectra with lower \epo\ will have a larger fraction of data points above
  the single power law fit and they will cause the fit with the single power
  law to be softer. This accounts for the observed positive correlation.  It
  is then dangerous to use this correlation for those cases where \epo\ is not
  measured, to derive it on the basis of $\alpha_{pl}$.
  
  Recently B07 derived the spectral properties of almost all the bursts
  detected by \sw\ through a Bayesian method. The method proposed in B07 to
  estimate the peak energy, far outside the energy range of BAT/\sw\ (15--150
  keV) is based on the assumption of a prior distribution for the observed
  peak energy which is the (gaussian) peak energy distribution of BATSE bright
  bursts (Kaneko et al. 2005), for \epo$>$300 keV, and a uniform distribution
  for energies below this value. However, as also noted by B07, from the
  comparison of these estimates of \epo\ with those derived from the spectral
  fits of the \ko\ and \su, for the common bursts, there could be a bias at
  high values of \epo\ estimated from \sw\ data through the Bayesian method
  and this could influence the findings of B07.

  In this paper, we adopt a conservative approach and we consider only the
  \sw\ GRBs for which the estimates of the peak energy is based on a standard
  (i.e.  frequentist) fit of the BAT spectra of bursts with known redshifts
  (from C07).

\section{No redshift evolution of the \ama\ correlation}

The 76 bursts of Tab. \ref{tab1} are distributed in a large redshift range, up
to $z=6.3$ (for GRB~050904 -- Tagliaferri et al. 2005).  It is worth to
investigate if any evolution with $z$ can affect the \ama\ correlation defined
by this sample.  Note that the simple evolution of the burst energetics
(\eiso) with $z$ does not necessarily implies the evolution of the \ama\ 
correlation.  Indeed, if \ep\ evolves in the same way as \eiso\ (i.e. if the
link between \eiso\ and \ep\ is physically robust) then we should see no
evolution of the slope and normalisation of the \ama\ correlation with $z$.

The possible redshift evolution of the \ama\ correlation has been investigated
by Li 2007 (L07, hereafter).  He used the sample of 48 GRBs (from Amati 2006,
2007), restricting to long events and excluding the peculiar GRB060614
(Gherels et al. 2006) and dividing the sample into 4 redshift bins.  Comparing
the slopes of the correlation corresponding to each redshift bin he found that
the \ama\ correlation becomes {\it steeper} for increasing redshift. Note that
the slope defined by L07 refers to the $E_{\rm iso} \propto E_{\rm
  peak}^{1/a}$ correlation, i.e. the reverse of that defined in this paper.
This is opposite to what one naively expects, namely that for larger $z$ we
select, on average, more energetic bursts with the same \ep, resulting in a
{\it flattening} of the \ama\ slope.  Note also from Fig.  \ref{amati} that
bursts with different redshifts are distributed along the \ama\ correlation
with no evident segregation along the correlation itself (except for the
excess of low redshift bursts in the low end of the \ama\ plane).

In our study we examine first the possible evolution with redshift of the
\ama\ correlation defined as $E_{\rm peak} \propto E_{\rm iso}^{a}$.  Our
sample of Tab. \ref{tab1} extends the original sample of 48 GRBs, used by L07,
both in number and redshift.  To verify his results, we have divided the 76
GRB sample into four redshift bins, chosen to have an equal number of 19
bursts per bin. These $z$ bins are very similar to those adopted by L07.  We
fit the \ama\ correlation in each sub--sample through the least square method
(see Sec. 2).  The results are shown in the insert of Fig. \ref{amati}.
The analysis of the \ama\ correlation in each redshift bin excludes the
evolution of the slope of the \ama\ correlation with $z$, given the present
sample of 76 events. Indeed, the slopes of the \ama\ correlations defined by
the four redshift sub--samples are consistent among themselves and also with
the slope of the correlation defined by the entire sample (insert of Fig.
\ref{amati}). We have also verified that the normalisation of the correlations
defined by the four redshift bins are consistent.

We then investigated the possible reasons for the discrepancy of our results
with respect to L07.  We reconstructed the sample of 48 GRBs used by L07 (with
data taken from Amati et al. 2006, 2007). Using this sub--sample we find the same
results of Li (2007), concerning the evolution of the slope and normalization
of the \ama\ correlation with redshift (circles in Fig. \ref{li}).  However,
when extending this sample to the 76 GRBs of our present sample, we do not
find any evolution with redshift (star symbols in Fig. \ref{li}). Note that
for this comparison we used the same fitting method of L07, which weights for
the errors on both variables, and fitted the $E_{\rm iso} \propto E_{\rm
  peak}^{1/a}$ correlation.  We then conclude that the Li (2007) results were
affected by too low statistics, and that there is no evidence that the \ama\ 
correlation evolves.

\setcounter{table}{0}
\begin{table*}
\begin{tabular}{lllllllll}
\hline 
GRB        &$z$    &$\alpha$        &$\beta$         &Fluence           &Range     &$E_{\rm peak}$ & $E_{\rm iso}$   & Ref \\ 
           &       &                &                &erg cm$^{-2}$     &keV       &keV            & erg             &     \\
\hline
970228     &0.695  &$-$1.54 [0.08]  &$-$2.5  [0.4  ] &1.1e--5  [0.1e--5]  &40--700    &195   [64]  &1.60e52  [0.12e52] & 1  \\        
970508$^a$ &0.835  &$-$1.71 [0.1 ]  &$-$2.2  [0.25 ] &1.8e--6  [0.3e--6]  &40--700    &145   [43]  &6.12e51  [1.3e51]  & 1 \\         
970828     &0.958  &$-$0.70 [0.08]  &$-$2.1  [0.4  ] &9.6e--5  [0.9e--5]  &20--2000   &583   [117] &2.96e53  [0.35e53] & 2  \\        
971214     &3.42   &$-$0.76 [0.1 ]  &$-$2.7  [1.1  ] &8.8e--6  [0.9e--6]  &40--700    &685   [133] &2.11e53  [0.24e53] & 1  \\        
980326     &1.0    &$-$1.23 [0.21]  &$-$2.48 [0.31 ] &7.5e--7  [1.5e--7]  &40--700    &71    [36]  &4.82e51  [0.86e51] & 1  \\        
980613     &1.096  &$-$1.43 [0.24]  &$-$2.7  [0.6  ] &1.e--6   [0.2e--6]  &40--700    &194   [89]  &5.9e51   [0.95e51] & 1  \\        
980703     &0.966  &$-$1.31 [0.14]  &$-$2.39 [0.26 ] &2.3e--5  [0.2e--5]  &20--2000   &499   [100] &6.90e52  [0.82e52] & 3  \\        
990123     &1.600  &$-$0.89 [0.08]  &$-$2.45 [0.97 ] &3.e--4   [0.4e--4]  &40--700    &2031  [161] &2.39e54  [0.28e54] & 1  \\        
990506     &1.30   &$-$1.37 [0.15]  &$-$2.15 [0.38 ] &1.9e--4  [0.2e--4]  &20--2000   &653   [130] &9.5e53   [1.13e53] & 2  \\        
990510     &1.619  &$-$1.23 [0.05]  &$-$2.7  [0.4  ] &1.9e--5  [0.2e--5]  &40--700    &423   [42]  &1.78e53  [0.26e53] & 1  \\        
990705     &0.843  &$-$1.05 [0.21]  &$-$2.2  [0.1  ] &7.5e--5  [0.8e--5]  &40--700    &348   [28]  &1.82e53  [0.23e53] & 1   \\       
990712     &0.433  &$-$1.88 [0.07]  &$-$2.48 [0.56 ] &6.5e--6  [0.3e--6]  &40--700    &93    [15]  &6.72e51  [1.29e51] & 1   \\       
991208     &0.706  &$ $...  ...     &$$....  ...     &1.6e--4  [5.0e--6]  &20--10000  &313   [31]  &2.23e53  [0.18e53] & 5   \\       
991216     &1.02   &$-$1.23 [0.13]  &$-$2.18 [0.39 ] &1.9e--4  [0.2e--4]  &20--2000   &642   [129] &6.75e53  [0.81e53] & 2   \\       
000131     &4.50   &$-$0.69 [0.08]  &$-$2.07 [0.37 ] &4.2e--5  [0.4e--5]  &20--2000   &714   [142] &1.84e54  [0.22e54] & 2   \\       
000210     &0.846  &$$...   ...     &$$....  ...     &7.6e--5  [5.0e--6]  &20--10000  &753   [26]  &1.49e53  [0.16e53] & 5  \\        
000418     &1.12   &$$...   ...     &$$....  ...     &2.6e--6  [4.0e--5]  &20--10000  &284   [21]  &0.91e53  [0.17e53] & 5  \\        
000911     &1.06   &$-$1.11 [0.12]  &$-$2.32 [0.41 ] &2.2e--4  [0.2e--4]  &15--8000   &1856  [371] &6.7e53   [1.4e53]  & 5  \\        
000926     &2.07   &$$...   ...     &$$....   ...    &2.6e--5  [4.e--6]   &20--2000   &310.  [20]  &2.7e53   [0.58e53] & 5  \\        
010222     &1.48   &$$...   ...     &$$....   ...    &1.4e--4  [8.e--6]   &20--2000   &766   [30]  &8.1e53   [0.86e52] & 5  \\        
010921     &0.45   &$-$1.60 [0.1 ]  &$$....  ...     &1.84e--5 [0.1e--5]  &2--400     &129.  [26]  &1.1e52   [0.11e52] & 5  \\        
011211$^b$ &2.140  &$-$0.84 [0.09]  &$$....  ...     &2.6e--6  [0.3e--6]  &40--700    &185   [25]  &6.64e52  [1.32e52] & (7) 2\\       
020124     &3.198  &$-$0.87 [0.17]  &$-$2.6  [0.65 ] &8.1e--6  [0.9e--6]  &2--400     &390   [113] &2.15e53  [0.73e53] & 8   \\       
020405     &0.695  &$-$0.00 [0.25]  &$-$1.87 [0.23 ] &7.4e--5  [0.7e--5]  &15--2000   &617   [171] &1.25e53  [0.13e53] & 2   \\       
020813$^b$ &1.255  &$-$1.05 [0.11]  &$$....  ...     &1.0e--4  [0.1e--4]  &30--400    &478   [95]  &6.77e53  [1.00e53] & 3   \\       
020819B    &0.41   &$-$0.90 [0.15]  &$-$2.0  [0.35 ] &8.8e--6  [9.e--7]   &2--400     &70.   [21]  &6.8e51   [0.17e51] & (7) 5\\       
020903$^b$ &0.25   &$-$1.00 [0.0 ]  &$$...   ...     &5.9e--8  [1.4e--8]  &2--10      &3.37  [1.79]&1.8d49   [0.31d49] & 9   \\       
021004$^b$ &2.335  &$-$1.00 [0.2 ]  &$$....  ...     &2.6e--6  [0.6e--6]  &2--400     &267   [117] &4.09e52  [0.71e52] & 7   \\       
021211     &1.01   &$-$0.85 [0.09]  &$-$2.37 [0.42 ] &2.2e--6  [0.2e--6]  &30--400    &94    [19]  &1.1e52   [0.13e52] & 2    \\      
030226$^b$ &1.986  &$-$0.90 [0.2 ]  &$$....  ...     &5.6e--6  [0.6e--6]  &2--400     &290   [63]  &6.7e52   [1.20e52] & 7     \\     
030328     &1.520  &$-$1.14 [0.03]  &$-$2.1  [0.3  ] &3.7e--5  [0.14e--5] &2--400     &328   [35]  &3.61e53  [0.40e53] & 3    \\      
030329     &0.169  &$-$1.32 [0.02]  &$-$2.44 [0.08 ] &1.2e--4  [0.12e--4] &3--400     &79    [3]   &1.66e52  [0.20e52] & 3    \\      
030429$^b$ &2.656  &$-$1.10 [0.3 ]  &$$...  ...      &8.5e--7  [1.4e--7]  &2--400     &128   [37]  &1.73e52  [0.31e52] & 7    \\      
040924$^c$ &0.859  &$$...   ...     &$$....  ...     &2.6e--6  [0.0]      &30--400    &102   [35]  &0.95e52  [0.1e52]  & 5    \\      
041006$^b$ &0.716  &$-$1.37 [0.14]  &$$...   ...     &2.0e--5  [0.2e--5]  &25--100    &108   [22]  &8.30e52  [1.3e52]  & 3    \\         
050126$^b$ &1.29   &$-0$.75 [0.44]  &$$...    ...    &8.55e--7 [1.82e--7] &15--150    &263   [110] &7.36e51  [1.60e51] & 10  \\       
050223$^{d,b}$&0.5915 &$-$1.50 [0.42]&$$...   ...    &6.14e--7 [0.83e--7] &15--150    &110   [54]  &1.21e51  [1.77e50] & 10  \\       
050318$^b$ &1.44   &$-$1.34 [0.32]  &$$...     ...   &2.1e--6  [0.2e--7]  &15--350    &115   [27]  &2.00e52  [0.31e52] & 11  \\       
050401$^e$ &2.9    &$-$1.00 [0.0 ]  &$-$2.45   ...   &1.93e--5 [0.04e--5] &20--2000   &501   [117] &4.1e53   [0.8e53]  & 13  \\       
050416A    &0.653  &$-$1.01 [0.0 ]  &$-$3.4    ...   &3.5e--7  [0.3e--7]  &15--150    &28.6  [8.3] &8.3e50   [2.9e50]  & 14  \\       
050505$^b$ &4.27   &$-$0.95 [0.31]  &$$...     ...   &2.58e--6 [3.06e--7] &15--150    &661   [245] &1.76e53  [2.61e52] & 10  \\       
050525A$^b$&0.606  &$-$0.01 [0.11]  &$$....    ...   &2.01e--5 [0.05e--5] &15--350    &127   [5.5] &2.89e52  [0.57e52] & 25  \\       
050603     &2.821  &$-$0.79 [0.06]  &$-$2.15 [0.09]  &3.41e--5 [0.06e--5] &20--3000   &1333  [107] &5.98e53  [0.4e53]  & 15  \\       
050803$^b$ &0.422  &$-$0.99 [0.37]  &$$...     ...   &2.08e--6 [2.57e--7] &15--150    &138   [48]  &1.86e51  [3.99e50] & 10  \\       
050814$^b$ &5.3    &$-$0.58 [0.56]  &$$...     ...   &1.46e--6 [1.16e--7] &15--150    &339   [47]  &1.12e53  [2.43e52] & 10  \\       
050820A$^b$&2.612  &$-$1.12 [0.14]  &$$...     ...   &5.27e--5 [1.2e--5]  &20--1000   &1325  [277] &9.75e53  [0.77e53] & 16  \\       
050904     &6.29   &$-$1.11 [0.06]  &$-$2.2  [0.4]   &5.4e--6  [1.e--7]   &15--150    &3178  [1094]&1.24e54  [0.1e54]  & 26  \\           
050908$^b$ &3.344  &$-$1.26 [0.48]  &$$...     ...   &4.36e--7 [0.46e--7] &15--150    &195   [36]  &1.97e52  [3.21e51] & 10  \\       
050922C$^b$&2.198  &$-$0.83 [0.26]  &$$....    ...   &2.6e--6  [0.26e--6] &30--400    &417   [118] &4.53e52  [0.78e52] & 17  \\       
051022$^b$ &0.80   &$-$1.17 [0.038] &$$....    ...   &2.61e--4 [0.8e--4]  &20--2000   &918   [63]  &5.3e53   [0.5e53]  & 27 \\         
051109A$^b$&2.346  &$-$1.25 [0.5 ]  &$$....    ...   &4.0e--6  [1.0e--6]  &20--500    &539   [381] &7.52e52  [0.88e52] & 18  \\      
060115$^b$ &3.53   &$-$1.13 [0.32]  &$$...     ...   &1.6e--6  [1.07e--7] &15--150    &288   [47]  &7.38e52  [1.23e52] & 10  \\      
060124$^b$ &2.297  &$-$1.48 [0.02]  &$$...     ...   &2.5e--5  [0.35e--5] &20--2000   &636   [162] &4.3e53   [0.34e53] & 19  \\      
060206$^b$ &4.048  &$-$1.06 [0.34]  &$$....    ...   &8.4e--7  [0.4e--7]  &15--150    &381   [98]  &4.68e52  [0.71e52] & 20  \\      
060210$^b$ &3.91   &$-$1.12 [0.26]  &$$...     ...   &6.92e--6 [3.74e--7] &15--150    &575   [186] &4.15e53  [5.7e52]  & 10  \\      
060218$^b$ &0.0331 &$-$1.62 [0.16]  &$$...     ...   &3.7e--6  [0.37e--6] &15--150    &4.9   [0.3] &7.7e49   [1.42e49] & 29 (31)\\     
060223A$^b$&4.41   &$-$1.16 [0.35]  &$$...     ...   &6.54e--7 [0.52e--7] &15--150    &339   [63]  &4.29e52  [6.64e51] & 10   \\     
060418$^b$ &1.489  &$-$1.50 [0.15]  &$$...     ...   &1.6e--5  [0.16e--5] &20--1100   &572   [114] &1.28e53  [0.10e53] & 21   \\     
060510B$^b$&4.9    &$-$1.53 [0.19]  &$$....    ...   &3.86e--6 [2.88e--7] &15--150    &575   [227] &3.67e53  [2.87e52] & 10   \\     
060522$^b$ &5.11   &$-$0.70 [0.44]  &$$....    ...   &1.05e--6 [1.04e--7] &15--150    &427   [79]  &7.77e52  [1.52e52] & 10   \\     
\hline
\end{tabular}
\caption{continue....}
\end{table*}

\setcounter{table}{0}
\begin{table*}
\begin{tabular}{lllllllll}
\hline 
GRB         & z     &$\alpha$      &$\beta$  &Fluence                    &Range   &$E_{\rm peak}$ &$E_{\rm iso}$   & Ref \\ 
            &       &              &         &erg cm$^{-2}$              &keV     &keV            &erg             &     \\
\hline
060526$^f$  &3.21   &$-$1.1  [0.4 ] &$-$2.2  [0.4] &4.9e--7   [0.6e--7]  &15--150   &105.25 [21.1]&2.58e52 [0.26e52] & 26    \\     
060605$^b$  &3.78   &$-$1.0  [0.44] &...     ...   &5.33e--7  [1.12e--7] &15--150   &490    [251] &2.83e52 [4.5e51]  & 10    \\    
060607A$^b$ &3.082  &$-$1.09 [0.19] &...     ...   &2.49e--6  [1.58e--7] &15--150   &575    [200] &1.09e53 [1.55e52] & 10    \\    
060614      &0.125  &$-$10   [0   ] &...     ...   &2.2e--5   [0.22e--5] &15--150   &55     [45]  &2.51e51 [1.0e51]  & 28    \\    
060707$^b$  &3.43   &$-$0.73 [0.4 ] &...     ...   &1.63e--6  [1.13e--7] &15--150   &302    [42]  &6.62e52 [1.39e52] & 10    \\    
060714$^b$  &2.711  &$-$1.77 [0.24] &...     ...   &3.06e--6  [3.96e--7] &15--150   &234    [109] &1.34e53 [9.12e51] & 10    \\    
060814      &0.84   &$-$1.43 [0.16] &...     ...   &2.69e--5  [0.26e--5] &20--1000  &473    [155] &7.01e52 [0.7e52]  & 32     \\
060904B$^b$ &0.703  &$-$1.07 [0.37] &...     ...   &1.48e--6  [1.91e--7] &15--150   &135    [41]  &3.64e51 [7.43e50] & 10    \\    
060906$^b$  &3.686  &$-$1.6  [0.31] &...     ...   &2.38e--6  [1.92e--7] &15--150   &209    [43]  &1.49e53 [1.56e52] & 10    \\    
060908$^b$  &2.43   &$-$0.9  [0.17] &...     ...   &2.66e--6  [1.11e--7] &15--150   &479    [110] &7.79e52 [1.35e52] & 10    \\    
060927$^b$  &5.6    &$-$0.93 [0.38] &...     ...   &1.1e--6   [0.1e--6]  &15--150   &473    [116] &9.55e52 [1.48e52] & 22    \\    
061007      &1.261  &$-$0.7  [0.04] &$-$2.61 [0.21]&2.5e--4   [0.15e--4] &20--10000 &902    [43]  &8.82e53 [0.98e53] & 23    \\   
061121$^b$  &1.314  &$-$1.32 [0.05] &....    ...   &5.7e--5   [0.4e--5]  &20--5000  &1289   [153] &2.61e53 [0.3e53]  & 24    \\   
061126$^b$  &1.1588 &$-$1.06 [0.07] &...     ...   &3.0e--5   [0.4e--5]  &15--2000  &1337   [410] &3.0e53  [0.3e53]  & 12    \\   
061222B$^b$ &3.355  &$-$1.3  [0.37] &...     ...   &2.24e--6  [1.23e--7] &15--150   &200    [28]  &1.03e53 [1.6e52]  & 10    \\  
070125      &1.547  &$-$1.1  [0.1 ] &$-$2.08 [0.13]&1.74e--4  [0.17e--4] &20--10000 &934    [148] &9.3e53  [0.93e53] & 30     \\ 
\hline
\end{tabular}
\caption{
$^a$ GRB 970508 is included following Amati (2006), but it was excluded in 
Ghirlanda et al. (2004) and Ghirlanda et al. (2007)
because of the inconsistency of the BATSE and $Beppo$SAX spectrum (i.e. it is a 
candidate outlier if the BATSE spectrum is assumed).  
In the sample of bursts before 2005 there are some \ep\ and \eiso\  values which are 
different from the recent compilation of Amati (2006). This is because he takes the average when two or 
more instruments (e.g. {\it Konus} and {\it Hete}) made the spectrum. 
This is correct when the spectral results are consistent, but we do not adopt the same
values if the spectral results are inconsistent.  
$^b$ \eiso\ has been computed as the logarithmic average of the values found 
with the Band (with $\beta$ fixed at --2.3) and cutoff power--law model 
(see also Ghirlanda et al. 2007, Firmani et al. 2005).
$^c$ We adopt the peak energy and isotropic energy values reported in Amati et al. (2006). 
There are only two GCNs for this burst giving the fluence and peak energy derived by
{\it Hete--II} and {\it Konus}, which are consistent. 
We then take the average of the two values.  
$\alpha$ values are not given in the GCN.
$^d$ The redshift is known but although it is present in B07 they do not report the redshift.
$^e$ We adopt an average of the spectral properties of the two peaks reported by the GCNs. 
Note that the two spectra are consistent.  
$^f$ This is the burst discussed on our webpage (http://www.brera.inaf.it/utenti/gabriele/060526.html).
This burst is an outlier of the Amati relation if the peak energy is much larger. 
References: before 2005:
(1) Amati et al. (2002);
(2) Ghirlanda et al. (2004, and references therein);
(3) Nava et al. (2006 and references therein);
(4) Ulanov et al. (2005);
(5) Amati (2006 and references therein);
(6) Firmani et al. (2006);
(7) Sakamoto et al. (2005);
(8) Atteia et al. (2005);
(9) Sakamoto et al. (2004);
Since 2005:
(10) Cabrera et al. (2007).
For the error the geometric mean is computed from the extremes of the logarithmic values);
(11) Perri et al. (2005);
(12) Perley et al. (2007, spectral analysis of the BAT+{\it RHESSI} spectrum); 
(13) Golenetskii et al. (2005);
(14) Sato et al. (2005);
(15) Golenetskii et al. (2005a);
(16) Cenko et al. (2006);
(17) Crew et al. (2005);
(18) Golenetstkii et al., (2005b);
(19) Romano et al. (2006);
(20) Palmer et al. (2006);
(21) Golenetskii et al. (2006);
(22) Stamatikos et al. (2006);
(23) Golenetskii et al. (2006a);
(24) Golenetskii et al. (2006b, \ep\ and \eiso\ are the average of the 
values found by the {\it RHESSI} and {\it Konus} spectra, while the spectral 
parameters are of the {\it Konus} spectrum);
(25) Blustin et al. (2005);
(26) Schaefer (2007);
(27) Golenetskii et al. (2006c);
(28) Mangano et al. (2007);
(29) Campana et al. (2006);
(30) Golenetskii et al. (2007). The burst was also seen by {\it RHESSI} (Bellm et al. 
2007) whose spectrum is inconsistent with that of {\it Konus}. However, the two
\ep\ and the fluences values given in the GCN are inconsistent with the average
\ep\ reported in the same GCN (for {\it RHESSI}). For this reason we report and use
the {\it Konus} spectral data; 
(31) Ghisellini et al. (2006);
(32) Golenetskii et al. (2006d).
}
\label{tab1}
\end{table*}

\section{The \epo--Fluence correlation}

The rest frame \ama\ correlation is defined by multiplying the
observed peak energy by $(1+z)$ and the observed (bolometric) fluence
by $4\pi d_{\rm L}^2/(1+z)$, where $d_{\rm L}$ is the luminosity
distance.  Therefore, in principle, the \ama\ correlation could be
only apparent, being the result of multiplying \epo\ and the fluence
$F$ by strong functions containing the redshifts (as argued by B07).

To study if this is the case, it is necessary to study what region of
the observational plane \epof\ is accessible to observations.  Any
detector, in fact, can only observe GRBs above a limiting fluence and
within a limited energy range.

Other selection effects could be related to the requirement of
measuring the redshift: for instance (in the absence of X--ray lines)
we require the burst to be located precisely enough to make possible
the optical follow up and we also require its afterglow to be visible
in the optical.  If only bright bursts had an optical afterglow, this
would affect the resulting correlation in the \ama\ plane.

We focus here on the selection effects introduced by requiring that
the burst is not only detected, but it must be bright enough to allow
the determination of the spectrum of its prompt emission.

Consider first how our 76 GRBs with known redshift are located
in the \epof\ plane, as illustrated in Fig. \ref{fl_ep_z.ps}.
Different symbols correspond to GRBs in different redshift bins.
Note that:
\begin{itemize}
\item These bursts define a strong correlation in this plane.  The Kendall's
  correlation coefficient is $\tau=0.53$ (9$\sigma$ significance). The least
  square fit has a slope 0.39$\pm$0.04. However, it is very likely that
    this correlation is affected by truncation effects which should be
    considered when computing the correlation strength and when recovering its
    ``true'' slope (LPM00).
\item There is no segregation in redshift.  We do not find that nearby bursts
  are brighter and bluer, and distant bursts are fainter and redder, as we
  would expect if there were no intrinsic correlation between \ep\ and \eiso.
\end{itemize}
%

\begin{figure}
\vskip -0.5 true cm
\centerline{\psfig{figure=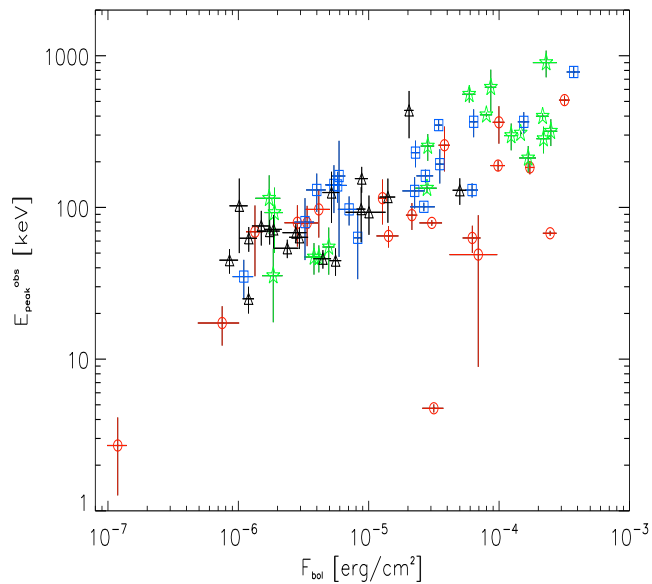,width=9.cm,height=8.5cm}} 
\vskip -0.3 true cm
\caption{
  \epo--bolometric fluence plane for the sample
   of 76 bursts of Tab. 1.  We show with different symbols the redshift bins
  corresponding to 0.03--0.843 (circles), 0.846--1.48 (stars), 1.49--3.1
  (squares) and 3.2--6.29 (triangles).  The least square fit
  to the data is  
  \epo$\propto F_{\rm bol}^{0.4}$ where $F_{\rm bol}$ is the
  fluence in the energy range 1 keV -- 10 MeV. The scatter of these bursts
  around the best fit has a $\sigma=0.21$.}
\label{fl_ep_z.ps}
\end{figure}

The existence of a \epof\ correlation is not new: it was discovered with
  a relatively small sample of \ba\ bursts by LPM00 and subsequently confirmed
  with the discovery of  the class of X--Ray Flashes by {\it Beppo}SAX and
{\it Hete--II} (e.g. Lamb, Donaghy \& Graziani 2005).  Sakamoto et al. (2005)
showed that a correlation between these two observables exists in the {\it
  Hete--II} burst sample and extends from the softest \epo\ of a few keV to
the few hundred keV range (e.g. Preece et al. 2000 and K06 for the BATSE
sample).  We will discuss the sample of bursts with no redshift, but with
measured \epo, in a forthcoming paper (Nava et al. 2008, in preparation).

\begin{figure*}
\vskip -0.5 true cm 
\centerline{\psfig{figure=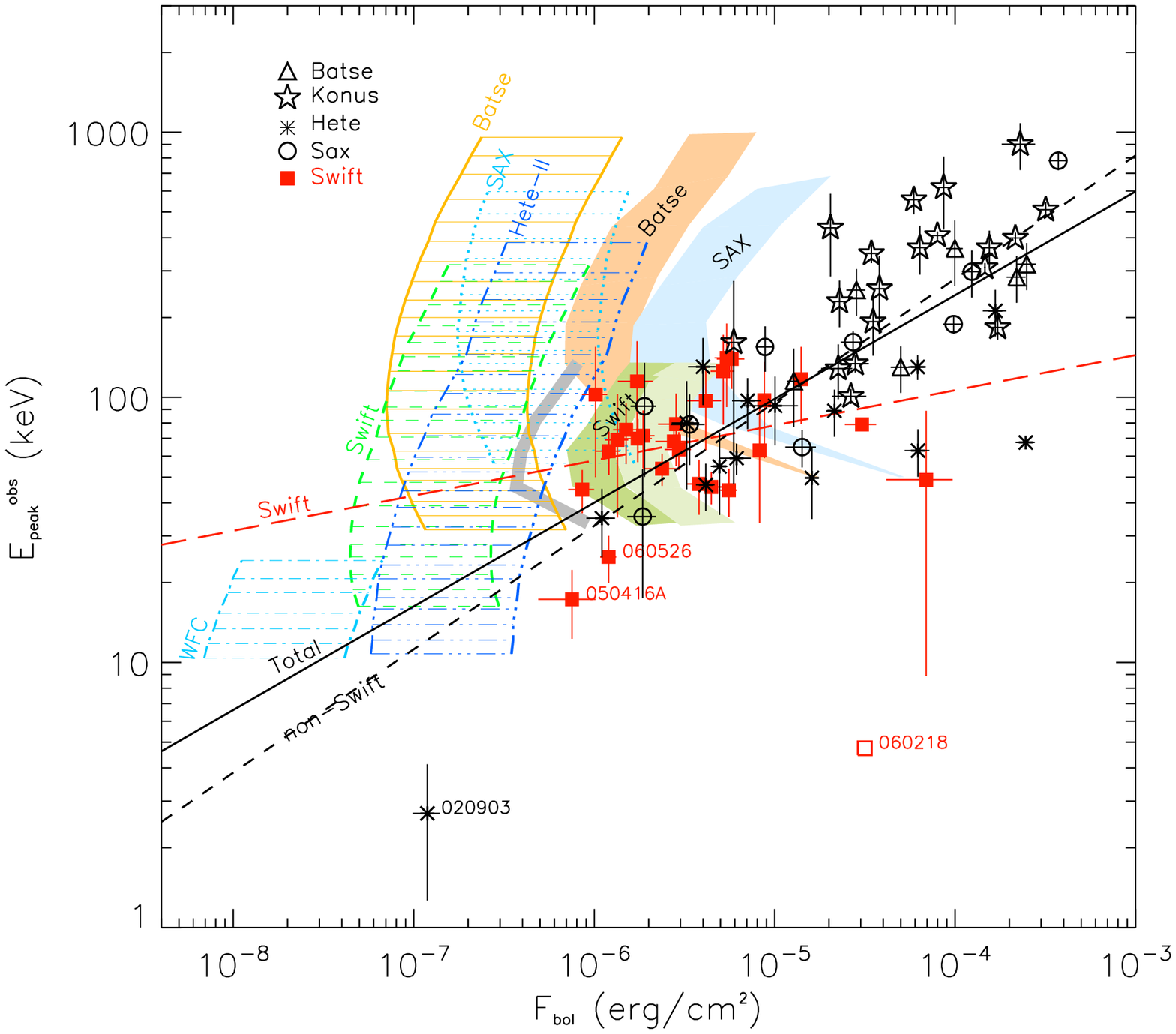,width=18cm,height=16cm}} 
\caption{ Observer frame plane of the peak energy and bolometric
  fluence (1--10$^4$ keV).  Bursts whose peak energy has been derived by the
  analysis of the BAT data are shown with filled squares. The open symbols are
  all the other bursts with redshift. The shades regions represent the minimum
  fluence that a burst can have (for different duration of the burst -- 5 s
  and 20 s for the left and right boundary of each shaded region,
  respectively) in order to perform the spectral analysis and derive
  constraints on \epo. For \sw\ we also report (narrow grey stripe) the
  limiting curve derived by slightly different criteria (see text for
  details).  The single lines represent the trigger sensitivity for \sw\ 
  (dashed), \ba\ (solid), \sax\ (dotted), \he\ (triple dot--dashed). The best
  fit of the sample of \sw\ bursts (long--dashed line) and non--\sw\ bursts
  (dashed line) and of the total sample (solid line) are also reported.  These
  fits, however, do not take into account the truncation effects (see text for
  details). The open red square represents GRB 060218 whose peak energy has
  been determined by the analysis of the \sw--XRT data (Campana et al. 2006).
}
\label{fluence_epicco}
\end{figure*}

\subsection{Instrumental biases in the \epof\ plane}

Consider the two regions in the \amaf\ plane of Fig. \ref{fl_ep_z.ps} below
and above the correlation where we have a paucity of points.  Region 1 is
characterized by large fluences and low/moderate \epo; region 2 comprises
small fluences, large/moderate \epo.  In region 1 there are no (or very few)
bursts, although there is no instrumental effect against their detection.
This implies that they really do not exist (or, rather, they are very few).
In region 2, instead, the paucity of points in the \amaf\ plane could be
affected by some instrumental selection effect.

The detectors (past and present) dedicated to study GRBs introduce at least
two selection effects on the population of bursts that they observe: (a) the
trigger sensitivity which is the ability to detect a burst as a transient
event significantly above the average background (see e.g. Band 2003; 2006);
(b) the ``spectral analysis threshold'' which is the minimum fluence required
to constrain the peak energy from the spectral analysis.

For any \epo, a {\it minimum} photon flux is required to trigger the burst.
This minimum flux depends on the spectral properties of the burst (especially
its peak energy -- e.g. Band 2003), on the detector design and on the total
background rate seen by the instrument.  The ability to detect a GRB for a
non--imaging instruments (like BATSE, {\it Konus, HeteII--Fregate},
$Beppo$SAX--GRBM), is related to the significance of the burst intensity (i.e.
its peak flux) with respect to an average background (see e.g. Band et al.
2003).  For imagers this threshold is modified by the coding aperture of the
mask and it can be even lower (Band 2006).  The trigger sensitivity curves as
a function of the burst peak spectral energy for different instruments has
been computed by Band (2003) (see also Band 2006).  We show them for \ba, \sw,
\he, \sax\ in Fig.  \ref{fluence_epicco}.  These curves correspond to the
sensitivity limit of each instrument to burst with a varying \epo and are
computed by Band (2003) in the peak flux -- peak energy space. In order to
report them on our \epof\ plane we have (i) to assume a typical burst duration
to cumpute the fluence and (ii) to assume a burst spectrum to convert from the
photon units to the energy units. To make this conversion we assumed a burst
with a 1 s duration which represents a lower limit for the detection sensivity
of long GRBs and we also assumed a burst with a typical Band spectrum (with
$\alpha=-1.0$ and $\beta=-2.3$) and with a variable \epo.  As discussed by
Band 2003, however, these curves are only an estimate of the burst detection
sensitivity for any given instrument because the trigger strongly depends on
the burst time profile and on the background.  In particular we tried to
  account for the unknown burst duration when converting the Band (2003)
  limiting curves in the fluence plane of Fig \ref{fluence_epicco}.  In the
  simplest case of a burst with a single pulse with a triangular shape, the
  conversion is $F(E_{\rm peak})=P(E_{\rm peak})T/2$ where $F(E_{\rm peak})$
  and $P(E_{\rm peak})$ are the fluence and peak flux as a function of the
  peak energy of the burst and $T$ is the burst duration.  Under this simple
  hypothesis, our curves for the trigger threshold would move in the plot of
  Fig. \ref{fluence_epicco} proportionally to $T/2$. However, GRBs light
  curves are all but simple single peaked triangles. A more realistic
  conversion factor between the peak flux and the fluence, which also accounts
  for the typical variation of the flux with time, can be obtained from the
  ratio of the fluence and the peak flux $F/P$ for the complete GRBs of the
  \ba\ sample (taking into account only the long burst population).  We found
  that the distribution of the ratio $F/P$ peaks at $\sim$6 and has a tail
  towards lower values.  In Fig. \ref{fluence_epicco} we report the two
  limiting curves for the trigger defined with a burst of duration 1 s and
  with the above ratio.  Note that the WFC trigger thresholds are on the left
  of the plot and are clearly not affecting the distribution of bursts with
  known redshifts. Moreover, these curves are here plotted in the very narrow
  energy range corresponding to the WFC trigger but their behaviour at higher
  \epo\ values has been shown to be very weak (see Fig. 3 of Band 2003). The
  WFC thresholds were taken from Band (2003) and they are consistent with what
  reported by Vetere et al. (2007)\footnote{Vetere et al. (2007) report a WFC
    detection threshold of $\sim 4\times 10^{-9}$ erg cm$^{-2}$ s$^{-1}$ in
    the 2--10 keV range which is in good agreement with the curves of Band
    (2003) when converted in bolometric flux (i.e. in the 1--10$^3$ keV
    band).}.  This is contrary to what one could think, i.e. that due to the
  requirement to accurately locate the burst in order to measure its redshift,
  it is the WFC trigger threshold to bias the GRBs (of the pre--\sw\ sample)
  with measured redshift in the \epof\ plane (but see also the discussion at
  the end of this section).  These results, taken at face values, imply that
  the distribution of GRBs of our heterogeneous sample in the \epof\ plane are
  not affected by the trigger sensitivity of the different istruments that
  detected these bursts.  

However, in order to perform the spectral analysis and to constrain any model
parameters we also require a {\it minimum} number of total (time integrated)
photons.  To derive the threshold fluence $F_{\rm min}$ required to perform
the spectral analysis, we have simulated several spectra with different \epo\ 
and fluence values.  For this simulation we need the detector response and a
typical background spectrum.  These two files are available for BATSE and \sw\ 
at the respective web sites.  Unfortunately, no public response file was found
for the {\it Beppo}SAX--GRBM instrument.  For the \sw\ simulations we used the
{\it batphasimerr} tool of the latest {\it HASOFTv6.3.2} release (C.
Markwardt, private communication).

We made two sets of simulations, assuming a burst durations of 5 and 20
seconds, respectively, with a corresponding different amount of background.

For each simulation we assumed an intrinsic spectrum described by the Band
model with typical spectral slopes (see e.g. K06) $\alpha=-1$ and
$\beta=-2.3$, a given peak energy and a given bolometric fluence.  The
resulting simulated spectrum is then fitted with a Band model (for BATSE
bursts) or with a cutoff--power law model (CPL, for \sw\ bursts) and with a
simple power law (PL).  The analysis returns the values of the spectral
parameters and the corresponding value of the reduced $\chi^2$, for both the
Band or CPL model and for the PL one.  We then repeat this procedure 500
times, constructing the distributions of the ``output" values of \epo\ (for
the Band and CPL models) and the associated error. We then fit the two
distributions with a Gaussian to find the average value of \epo, the average
value of the error with the associated $\sigma_{\rm E}$ value. The average
value of the ``output" \epo\ is consistent with the assumed \epo\ value.  We
then decide if the fitting is returning an acceptable value of \epo\ adopting
two criteria:
\begin{itemize}
\item
The error on \epo is less than 100\% in the majority (97.7\%)
of the 500 simulations.
\item
The fit with the Band or CPL models are significantly better than 
the fit with a single power law.
This is measured by the $F$--test, and we set the threshold to
2$\sigma$ (corresponding to a probability of 95.45\%).
\end{itemize}
Note that for \sw\ bursts we do not ask that an intrinsic spectrum following
the Band function can be better fitted with the same model, requiring only
that we can reconstruct \epo, even if only with a CPL model.  If both these
conditions are fulfilled, we repeat the set of simulations with a smaller
fluence and the same \epo, until one or both the conditions are not met.  This
defines the minimum fluence (for a given \epo).  This is illustrated in Fig.
\ref{fluence_epicco} by the shaded regions (they are not lines because they
correspond to two different duration times).  The values reported in Fig.
\ref{fluence_epicco} are the ``output" values of \epo\ and bolometric fluence,
since we want to compare them with the GRBs actually detected.  Our
simulations have been performed assuming a peak energy between 30 and 140 keV
for \sw\ and 50 and 1000 keV for \ba. For values of \ep\ outside this range it
is difficult to fit a cutoff--power law or Band spectrum due to the few
spectral bins between the peak energy and the end of the energy range.

In another set of simulations, performed only for \sw\ bursts, we relaxed the
above criteria, replacing them by the condition that the ``output value" of
\epo\ determined by a cut off power law model, and its associated 1$\sigma$
error, is contained in the 15--200 keV band. In other words, \epo$+\sigma_E$
has to be less then 200 keV, and \epo$-\sigma_E$ must be larger than 15 keV.
For these simulations we performed the same spectral fitting procedure
described in Section 2 of C07 (in particular a logarithmic spacing of \epo\ 
was used).  The resulting limiting curve is shown by the grey narrow stripe in
Fig. \ref{fluence_epicco} (corresponding to burst durations of 5 seconds).

Note that the simulations assume a burst observed on--axis for both \sw\ and
\ba.  While in the case of \ba\ the LAD sensitivity does not strongly change
for off--axis incidence, in the case of \sw--BAT the sensitivity strongly
depends on the photon incidence angle. For this reason we also considered for
\sw\ the case of a 60 deg off--axis burst (light shaded region in Fig.
\ref{fluence_epicco})\footnote{This is controlled by the {\it pcoderf} keyword
  of the {\it batphasimerr} tool}.  We also tested the dependence of the
simulations from the assumed shape of the spectrum: we repeated the
simulations with $\alpha=-0.5$ (i.e. in the tail of the distribution of this
parameter for the \ba\ bright burst population -- K06) and found that the
limiting curves in Fig. \ref{fluence_epicco} move to the left by a factor 2 on
average.

To understand the shape of the limiting region in Fig. \ref{fluence_epicco} we
can make a simple argument.  It is reasonable to assume that the determination
of \epo\ (within a given confidence level) requires a minimum number of
photons $N_{\rm min}$ around the peak of the spectrum.  This $N_{\rm min}$
does not depend on the value of \epo, and corresponds to a minimum fluence
$F_{\rm min}$.  We have:
\begin{equation}
F_{\rm min}\, =\, N_{\rm min}\, 
{ E^{\rm obs}_{\rm peak} \over  A_{\rm eff}(E^{\rm obs}_{\rm peak}) }
\label{smin}
\end{equation}
where $A_{\rm eff}$ is the effective area of the detector.  Inverting Eq.
\ref{smin} we have a threshold curve \epo$(F_{\rm min})$: only for bursts with
fluence larger than this curve we can derive \epo.  For a constant $A_{\rm
  eff}$, bursts with larger \epo\ have larger $F_{\rm min}$ (since $N_{\rm
  min}$ is the same), and this in principle might explain why only few faint
bursts have large \epo.  For a fixed $N_{\rm min}$ and constant $A_{\rm eff}$,
the limiting curve in the \amaf\ plane is linear.  This is indeed very similar
to the shape of the limiting regions for \ba, \sw, and {\it Beppo}SAX--GRBM
reported in Fig. \ref{fluence_epicco}.

In Fig. \ref{fluence_epicco} we also show the limiting region for the spectral
analysis for {\it Beppo}SAX--GRBM. Although the response files of this
instrument has not been yet released, the GRBM and BATSE are similar
scintillators and their effective area are similar in the $\sim$20--800 keV
range. For this reason we can have an approximate idea of the limiting fluence
by rescaling the limiting \epo$(F_{\rm min})$ curves of BATSE for the ratio of
the effective areas of BATSE and GRBM.

We note that the ``spectral threshold'' is not affecting the sample of \ba\ 
and \sax\ bursts (solid open symbols in Fig. \ref{fluence_epicco}).  It is
therefore compelling to investigate if there exists bursts detected by \ba\ or
\sax\ which lie between the limiting threshold curves and the sample of GRBs
with measured redshifts shown in Fig. \ref{fluence_epicco}.  These bursts
would not have a measured redshift and would not be affected by the selection
effect (if any) intervening when asking that GRB has its redshift measured.
In order to investigate these issues, however, it is necessary to have a
complete GRB sample down to the limiting fluence represented by the curves of
Fig. \ref{fluence_epicco} and we leave this to a forthcoming paper (Nava et
al. 2008).

We can see that the ``spectral threshold'' is affecting the \sw\ bursts
(filled red squares in Fig. \ref{fluence_epicco}): due to the the limited BAT
energy range, \sw\ can only add bursts in the $\sim$ 15--150 keV range of the
\epof\ plane.  Note that there are 3 \sw\ bursts which are below the spectral
thresholds defined above.  One of these is GRB 060218 (open square in Fig.
\ref{fluence_epicco}) whose spectral peak energy has been found by combining
the BAT and XRT \sw\ data (Campana et al. 2006). The other two are GRB 050416A
and GRB 060526. For GRB 050416A we adopted the spectral results reported in
Sato et al. (2007) which were derived through a fit with a Band model with a
fixed low energy spectral index in order to account for the softness of the
burst (see also Sakamoto et al. 2006).  The spectrum of this burst can be well
fitted also with a single power law softer than $\Gamma=2$. For GRB 060526 we
adopted the spectral parameters reported in Schaefer et al. (2006).  However,
our analysis of the BAT
data\footnote{www.brera.inaf.it/utenti/gabriele/060526.html} were consistent
also with a single power law with $\Gamma\sim1.7$.

\begin{figure}
  \vskip -0.5 true cm 
\centerline{\psfig{figure=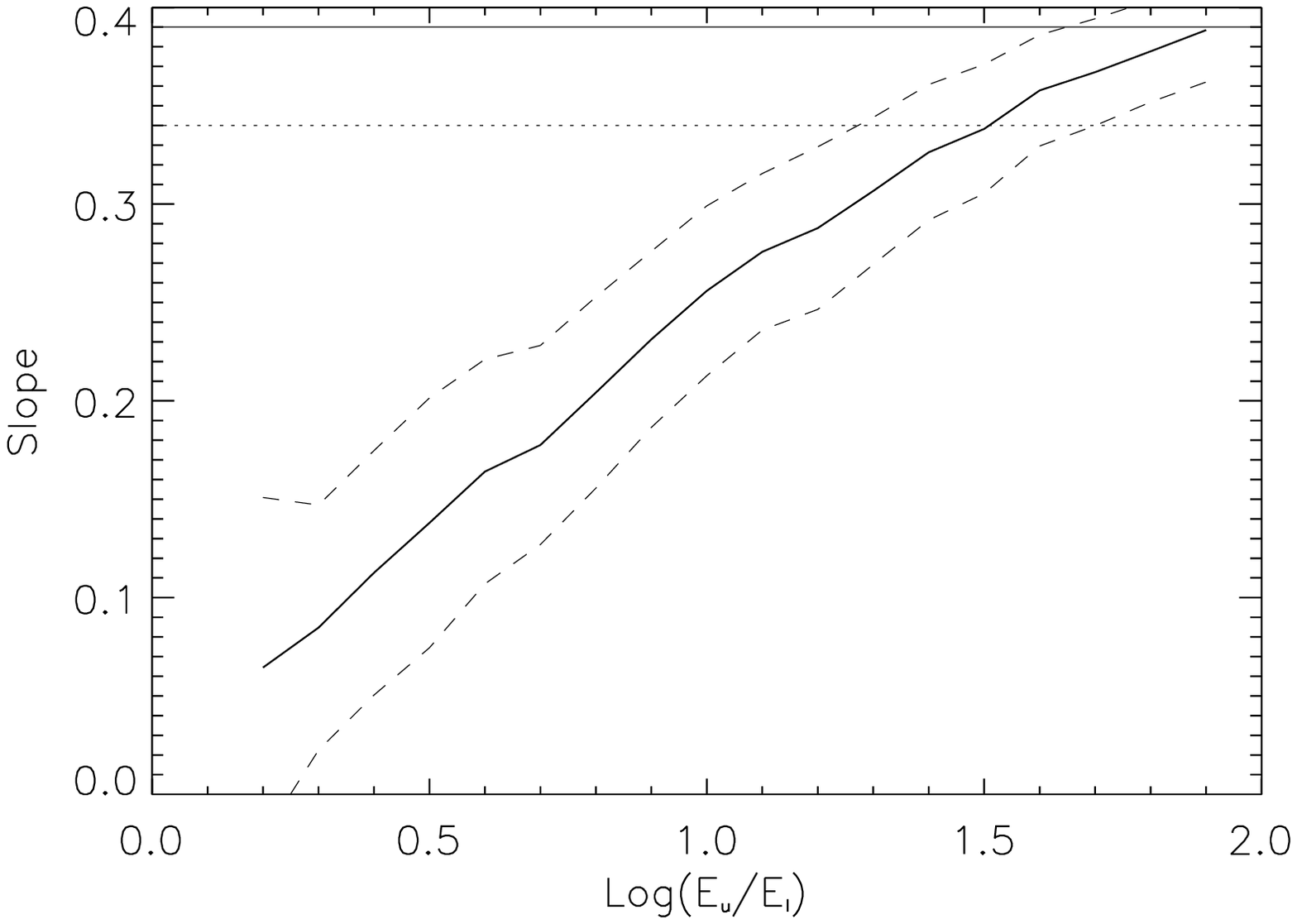,width=9.cm,height=8.5cm}} 
\caption{Simulation of the dependence of the slope of the \epof\ correlation 
  on the energy interval width $\log(E_{\rm u}/E_{\rm l})$ in \epo\ used to
  compute it. Here $E_{\rm u}$ and $E_{\rm l}$ represent the boundary of the
  interval in \epo. The original correlation (with slope 0.39) and its 90\%
  confidence level are shown by the solid and dotted orizontal lines,
  respectively. The solid line and the dashed lines represent the slope of the
  correlation obtained by selecting a sub--sample of bursts with peak energy
  comprised within a range of width $\log(E_{\rm u}/E_{\rm l})$. }
\label{simula}
\end{figure}

Having shown that at least the low end of the correlation in the \epof\ is
affected by a truncation effect, acting mainly on the \sw\ burst population,
we proceed to analyze the correlation following the method proposed by LPM00.
This method, however, can be applied only to the sub-samples of bursts for
  which the dominant truncation effect is known. We have shown that nor the
  trigger detection threshold neither the spectral analysis threshold of the
  different instruments that contributed to the heterogeneous GRB sample are
  affecting the GRBs with redshifts except for the \sw\ sample. We therefore
  can apply the correlation analysis on the \sw\ sub--sample taking into
  account its spectral analysis truncation effect.  
 
We calculate the
Kendall's correlation coefficient with only those \sw\ pairs of bursts which
lie above each other's spectral analysis thresholds.  We obtained $\tau=0.1$
(0.6$\sigma$ significant). The significance has been computed with the
boostrap method described in the Appendix of LPM00. This result suggests that
there is no \epof\ correlation in the \sw\ GRB sample considered.
 However, we stress that the \sw\ bursts suffer from another strong
  truncation effect: the bursts peak energy can be computed only if it lies in
  the 15--150 keV energy range where the spectra are fitted (but see B07 for a
  different approach). This is clearly shown in Fig. \ref{fluence_epicco} by
  the clustering of the \sw\ bursts (red squares) in this limited energy
  range. This energy range is smaller than the dispersion of the \epof\ 
  correlation defined by the pre--\sw\ sample and this implies that it cannot
  be used to probe the existence of the \epof\ correlation. Indeed, the
  dispersion of the \sw\ bursts along the fluence axis is much larger than the
  dispersion along the \epo\ axis only due to the limited energy range where
  the \sw\ burst peak energy is computed. 
  To prove this we performed a
  simulation: assume that the correlation defined by the 76 GRBs in the \epof\ 
  plane is true and extract randomly a subsample of bursts with \epo\ within
  the range $[E_{\rm l}$--$E_{\rm u}]$.
  We vary  the width of this range and compute the correlation of the resulting subsample.  
  The result is shown in Fig. \ref{simula}. 
  Note that the energy range in \epo\ of \sw\ bursts (red
  squares in Fig. \ref{fluence_epicco}, excluding 050416A, 060526 and 060218 -
  see above) corresponds to $E_{\rm u}/E_{\rm l}\sim 5$. 
  With this small energy range we do not recover the true correlation.  
  Only if $E_{\rm u}/E_{\rm l}\gsim 25$ we have a correlation 
  consistent with the simulated one.
  Based on this result, we conclude that while the \sw\ bursts alone do not
  show the \epof\ correlation shown by the pre--\sw\ 
  bursts, this sample cannot be used to rule out the existence of the very
  same correlation due to its limited range of \epo.  Considering instead all
  the other bursts we can conclude that the selection effects considered are
  not affecting their distribution in the \epof\ plane but still we cannot
  exlcude that other, unknown and still to be studied, selection effects could
  be present. In this paper we studied two of the most relevant selection
  effects, i.e. the condition of detecting a burst and that (more
  constraining) of properly fitting its spectrum to derive \epo\ and its
  fluence.

  Finally, there could be another selection effect responsible for the fact
  that the bursts of measured redshift have on average larger fluences. 
This selection effect could be associated to the requirement of 
deriving their position with an accuracy high
  enough to start the ground based follow--up. Indeed, while in the \sw\ 
  era the X--ray afterglow position is known with a typical few arcsec
  accuracy, in the pre--\sw\ era the positioning of the X--ray
  transient was given by the WFC  and WXM. In order to
  test this possibility we considered the sample of bursts detected by the WFC
  and GRBM on board \sax\ (Frontera 2004). In this case, as shown by the
  trigger threshold reported in Fig. \ref{fluence_epicco}, it was the GRBM
  trigger threshold to dominate over the WFC one. However, by comparing the
  distribution of the fluence of the sub--sample of \sax\ bursts with and
  without redshifts (Fig. \ref{sax}), we see that there is no preference for
  bursts with redshifts to have larger fluences
  (the Kolmogorov--Smirnov probability test is 0.38). 
  The bursts with redshift are about half of those
  detected by both the GRBM and WFC onboard \sax.
  This is expected, due to the fraction of dark bursts, and the visibility
  constraints of the ground based optical telescopes.
  
This result, regarding the \sax\ GRB sample, is puzzling.
In fact, in Fig. \ref{fluence_epicco}, the WFC limiting curves
  are a factor 100 on the left side of the distribution of the corresponding
  \sax\ bursts (also shown in Fig. \ref{sax}).
  We would have expected that the GRBs detected by both
  the WFC and the GRBM, without requiring the redshift to be measured,
  had minimum fluences corresponding to the limit posed by the GRBM.
  Instead they all lie at larger fluences, and
  we did not find any reason to explain this puzzle.
%

\begin{figure}
  \vskip -0.5 true cm 
\centerline{\psfig{figure=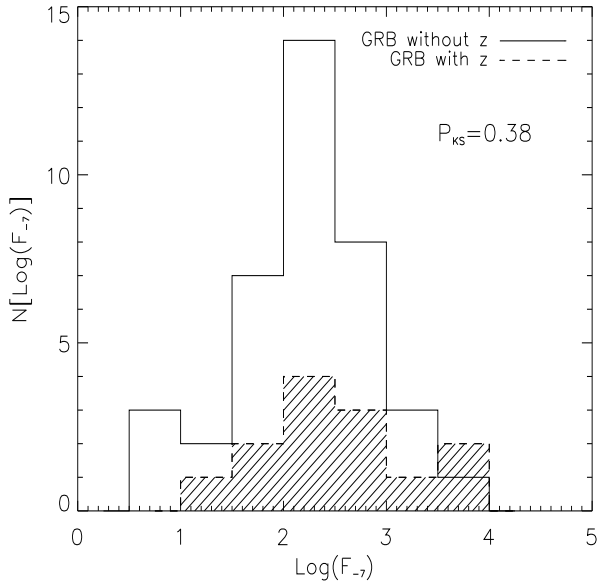,width=9.cm,height=8.5cm}} 
\caption{Distribution of the observed fluence (2--700 keV) of bursts detected
  by both the GRBM and the WFC on board \sax\ (Frontera 2004). The two
  histograms correspond to GRBs with (hatched) and without (empty) redshifts.
  The two distribution Kolmogorov--Smirnov test probability is 0.38. }
\label{sax}
\end{figure}


\begin{figure}
  \vskip -0.5 true cm 
\centerline{\psfig{figure=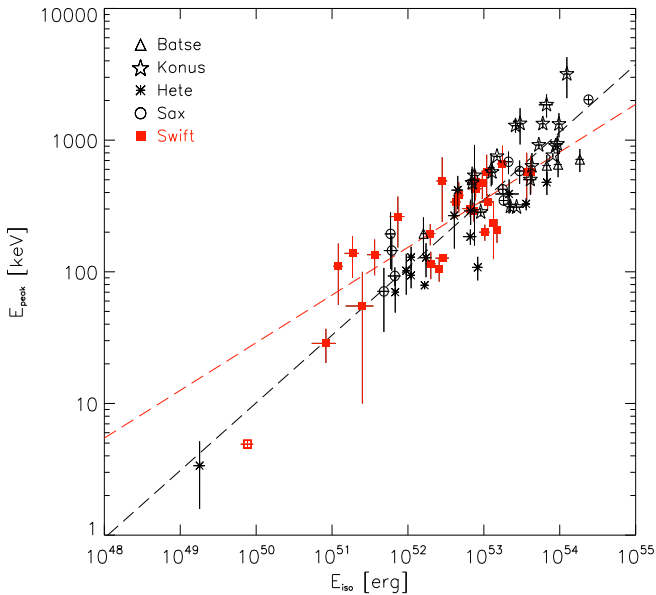,width=9.cm,height=8.5cm}} 
\caption{Rest frame \ama\ correlation. The \sw\ bursts whose peak energy has
  been derived from the analysis of the BAT data are shown with filled red
  squares (their fit -- excluding GRB 060218 -- is represented by the solid red
  line). The non--\sw\ bursts are shown by the open symbols and their fit by
  the long--dashed black line.}
\label{amati_swift}
\end{figure}


In Fig. \ref{amati_swift} we show in the \ama\ plane, separately, the
pre--\sw\ bursts and those whose spectral parameters have been determined only
by \sw\ data.  The best least square fit to the two sets of data yields a
slope $a=0.51\pm0.03$ for the pre--\sw\ GRBs and $a=0.36\pm0.05$ for the \sw\ 
bursts.  The two slopes become consistent if one performs a fit that weights
for the errors on both quantities, as found by C07. However, for the \sw\ 
  bursts of our sample, we have found the dominant selection effect. By
  accounting for this truncation we find a very weak correlation in the rest
  frame ($a=0.1$, Kendall's $\tau=0.12$ at 0.58$\sigma$).

\section{Summary and conclusions}

In this paper we have studied the \ama\ correlation defined with the most
updated sample (up to Sep. 2007) of GRBs with measured redshift and well
defined spectral properties: 35 GRBs detected before the launch of the \sw\ 
satellite (Nov. 2004) and 41 events detected since then. The latter sample is
composed by 27 events whose spectral properties were obtained through the
analysis of the \sw-BAT spectra and 14 bursts (in most cases also detected by
\sw) but with a spectrum detected by other satellites ({\it Konus, Suzaku,
  RHESSI}) with a larger spectral energy window. With this large GRB sample we
have studied the possible evolution with redshift, the existence of other
correlations among the spectral parameters or with redshift, the possible
instrumental selection effects acting on this sample of 76 GRBs with measured
redshift.

Our main results show that in the rest frame \ama\ plane the 76 GRBs define a
strong correlation with no new outlier (with respect to the classical GRB
980425 and GRB 031203 (see e.g. Ghisellini et al. 2006).  We find no
correlation between the spectral parameters and the redshift nor between the
peak energy and the spectral photon index.

By dividing the GRB sample into redshift bins we compared the slope of the
\ep\ correlation for each redshift bin. We do not find any evidence that this
correlation evolves with redshift (as also found by B07) contrary to what has
been found with a smaller GRB sample (Li 2007).

The analysis of the observer frame \epof\ plane shows the existence of
a strong correlation between the observed peak energy and the
fluence. This correlation is of the form
\epo$\propto$(Fluence)$^{0.4}$ and GRBs of different redshift are
evenly distributed along it.  The \epof\ plane is where the possible
selection effects on the \ep\ correlation should be investigated.  Two
regions of this plane are defined by the \epof\ correlation: bursts
with large fluence and small/intermediate peak energy should not exist
(or be very rare) as nothing prevents their detection, while bursts of
intermediate/large \epo\ and small fluence could be affected by some
instrumental selection effect.  We consider two of these: (a) the
minimum flux to trigger a burst and (b) the minimum fluence required
to fit its spectrum and constrain its peak energy.  We have used the
threshold curves derived by Band (2003) to represent the first
selection effect in the \epof\ plane.  We have converted these curves
into the fluence--\epo\ plane, assuming a range of burst durations.
We found that the trigger threshold does not affect the distribution
of bursts with known redshifts in the \epof\ plane. The second
selection effect that we considered (to our knowledge this is the
first time that such a selection effect is considered for the GRB
detectors) is related to the requirement to have enough photons in the
spectrum to fit it with a curved spectral model (either the Band model
-- Band et al. 1996 -- or a cutoff--power law model) and constrain the
peak energy of the burst. To this aim we performed detailed spectral
simulations which account for the detector response and the typical
GRB background.  We also accounted for some parameters (burst duration
and GRB off--axis position) which contribute to determine these
threshold curves. Our results (shown in Fig. \ref{fluence_epicco})
demonstrate that the latter selection effect dominates the first in
the \epof\ plane as shown in Fig.  \ref{fluence_epicco}.  Our results
show that the pre--\sw\ GRB sample, containing a fraction of bursts
detected by \sax\ and \ba\, is not affected by the corresponding
limiting curves, while the \sw\ sample is. Indeed, in the case
of \sw\ most of the GRB whose spectra are {\it well fitted} by a
cutoff--power law model have an \epo\ which is in the range 15--150
keV. The correlation in the \epof\ plane defined by the \sw\ sample is
flat and very weak compared to that defined with all the other bursts.
However, we cautiously note that the \sw\ sample is distributed in a
very narrow range in \epo,
smaller than the scatter of the \epof\ correlation as defined by all the
pre--\sw\ bursts.  As a consequence, although \sw\ bursts do not show a \epof\ 
correlation, we cannot rule out the existence of the very same correlation
defined by the non--\sw\ bursts, extending over 2 orders of magnitudes in both
\epo\ and Fluence.

Our results do not exclude that other selection effects affect the \epof\ 
plane, and in particular the non-\sw\ bursts.  The lack of many bursts
with known redshift with intermediate/large \epo and small fluence (i.e.
between the present sample of GRBs and the \ba\ limiting curves of Fig.
\ref{fluence_epicco}) could be due to the additional presence of a still not
understood selection effect, different from those considered in the present
paper (see also Nakar \& Piran 2005).  The first step to investigate this
possibility is to search if there exist GRBs which populate the region on the
left side of the \epof\ correlation. This issue cannot be solved by including,
in the \epof\ plot, the \ba\ bursts already spectrally analyzed by Kaneko et
al.  (2006), since they all have large fluences.  Furthermore, the sample of
Yonetoku et al.  (2004) could be biased by their requirement of having a
pseudo--redshift not exceeding 12, as measured by the \ep--luminosity
relation.  Therefore we need a new, representative sample of \ba\ bursts with
measured \epo\ and fluences close to the limiting fluence curve (see again
Fig.  \ref{fluence_epicco}).  This is what we will present in a forthcoming
paper (Nava et al. 2008, in preparation).  We anticipate here that these
bursts exist and therefore the conclusion that can be drawn from the present
work is that other selection effects, different from the the GRB trigger
threshold or the spectral threshold considered in the present paper, are very
likely affecting the \epof\ correlation and, as a consequence, the rest frame
\ama\ correlation.

\section*{Acknowledgements} 
We thank partial funding by a 2005 PRIN--INAF grant.  We thank ASI (I/088/06/0)
for funding.  We would like to thank the referee for her/his constructive
comments which helped to revise and improve the manuscript.


\begin{thebibliography}{}

\bibitem[]{} Amati, L., Frontera, F., Tavani, M., et al. 2002, A\&A, 390, 81
\bibitem[]{} Amati, L., 2006, MNRAS, 372, 233
\bibitem[]{} Atteia J.-L., 2005, NCimC, 28, 647
\bibitem[]{} Band, D. L. et al., 1993, ApJ, 413, 281
\bibitem[]{} Band, D. L., 2003, ApJ, 588, 945
\bibitem[]{} Band, D. L., 2006, ApJ, 644, 378
\bibitem[]{} Band, D.L. \& Preece, R.D., 2005, ApJ, 627, 319 
\bibitem[]{} Bellm, E. C., Hurley, K., Pal\'shin, V., 2007, arXiv:0710.4590
\bibitem[]{} Blustin, A. J., Band, D., Barhelmy, S., 2006, ApJ, 637, 901
\bibitem[]{} Bosnjak, Z., Celotti, A., Longo, F., Barbiellini, G., 2007, MNRAS
  subm., astro-ph/0502185
\bibitem[]{} Butler, N.R., Kocevski, D., Bloom, J.S. \& curtis, J.L., 2007, arXiv:0706.1275
\bibitem[]{} Cabrera, I.,   2007, MNRAS, in press (C07), arXiv:0704.0791
\bibitem[]{} Campana, S., Mangano, V., Blustin, A. J., et al., 2006, Nat.,
  442, 1008
\bibitem[]{} Cenko, S. B., Kasliwal, M., Harrison, F. A., 2006, ApJ, 652, 490
\bibitem[]{} Crew G., Ricker, G., Atteia, J.-L., 2005, GCN 4021
\bibitem[]{} D'Agostini, G., 2005, arXiv:physics/0511182
\bibitem[]{} Eichler, D. \& Levinson, A., 2004, ApJ, 614, L13
\bibitem[]{} Firmani C., Ghisellini, G., Avila-Reese, V. et al., 2006, MNRAS, 370, 185
\bibitem[]{} Frontera F., AIP conf proc. (astro-ph/ 0407633) 
\bibitem[]{} Gehrels, N., Chincarini, G., Giommi P., et al., 2004, ApJ, 611, 1005
\bibitem[]{} Ghirlanda, G., Ghisellini, G. \& Lazzati, D. 2004, ApJ, 616, 331
\bibitem[]{} Ghirlanda, G., Ghisellini, Firmani C., 2005, MNRAS, 361, 10L
\bibitem[]{} Ghirlanda, G, Nava, L., Ghisellini G., Firmani C., 2007, A\&A, 466, 127
\bibitem[]{} Ghisellini, G., Ghirlanda, G., Mereghetti, S., Bosnjak, Z., Tavecchio, F., 
             \& Firmani, C., 2006, MNRAS, 372, 1699 
\bibitem[]{} Golenetskii S., Aptekar R., Mazets E., et al., 2005, GCN 3179
\bibitem[]{} Golenetskii S., Aptekar R., Mazets E., et al., 2005a, GCN 3518
\bibitem[]{} Golenetskii S., Aptekar R., Mazets E., et al., 2005b, GCN 4328
\bibitem[]{} Golenetskii S., Aptekar R., Mazets E., et al., 2006, GCN 4989
\bibitem[]{} Golenetskii S., Aptekar R., Mazets E., et al., 2006a, GCN 5722
\bibitem[]{} Golenetskii S., Aptekar R., Mazets E., et al., 2006b, GCN 5837
\bibitem[]{} Golenetskii S., Aptekar R., Mazets E., et al., 2006c, GCN 4150
\bibitem[]{} Golenetskii S., Aptekar R., Mazets E., et al., 2006d, GCN 5460
\bibitem[]{} Golenetskii S., Aptekar R., Mazets E., et al., 2007, GCN 6049
\bibitem[]{} Guidorzi, C., Frontera, F., Montanari E., et al., 2005, MNRAS,
  363, 315
\bibitem[]{} Kaneko, Y., Preece, R.D., Briggs, M.S., Paciesas, W.S., Meegan, C.A. 
             \& Band, L., 2006, ApJS, 166, 298 (K06)
\bibitem[]{} Lamb, D.Q., Donaghy, T.Q. \& Graziani, C., 2005, ApJ, 620, 355 
\bibitem[]{} Li, L.--X., 2007, MNRAS, 379, L55 (L07)
\bibitem[]{} Liang, E. \& Zhang, B., 2005, ApJ, 633, L611  
\bibitem[]{} Lloyd--Ronning, N. \& Ramirez--Ruiz, E., 2002, ApJ, 576, 101L 
\bibitem[]{} Lloyd--Ronning, N., Petrosian, V. \& Mallozzi, R. S., 2000, ApJ,
  534, 227 (LPM00) 
\bibitem[]{} Mangano, V., Holland, S. T., Malesani, D., 2007, A\&A, 470, 105
\bibitem[]{} Nakar, E. \& Piran, T., 2005, MNRAS, 360, L73
\bibitem[]{} Nava, L., Ghisellini, G., Ghirlanda, G., Tavecchio, F. \& Firmani, C.
             2006, A\&A, 450, 471
\bibitem[]{} Nava, L., Ghisellini, G., Ghirlanda, G., Cabrera, J.I., Firmani, C. 
             \& Avila-Reese, V., 2007, MNRAS, 377, 1464
\bibitem[]{} Norris,  J. P., Marani, G. F., Bonnell, J. T., 2000, ApJ, 534, 248
\bibitem[]{} Norris,  J. P., 2002, ApJ, 579, 386
\bibitem[]{} Palmer, D., Barbier, L., Barthelmy, S., 2006, GCN 4697
\bibitem[]{} Perri, M., Giommi, P., Capalbi M., et al., 2005, A\&A, 442, L1
\bibitem[]{} Perley, D. A., Bloom, J. S., Butler, N. R., et al., 2007, ApJ, subm., astro-ph/0703538
\bibitem[]{} Preece, R. D., Briggs, M. S., Mallozzi, R. S., et al., 2000, ApJS, 126, 19
\bibitem[]{} Reichart D., Lamb D. Q., Fenimore E. E. et al., 2001, ApJ, 552,
  57
\bibitem[]{} Reichart D., astro-ph/0508529 
\bibitem[]{} Ramirez--Ruiz, E., Granot, J., Kouveliotou, C., Woosley, S.E., Patel, S.K. \& Mazzali, P.A.,
             2005, ApJ, 625, L91
\bibitem[]{} Romano P., Campana, S., Chincarini, G., 2006, A\&A, 456, 917
\bibitem[]{} Sakamoto, T., Nakagawa, Y., Torii, K., et al., 2004, AIPC, 727, 106
\bibitem[]{} Sakamoto, T., Lamb, D.Q., Kawai, N., et al., 2005, ApJ, 629, 311  
\bibitem[]{} Sakamoto, T., Sato, G., Barbier, L., et al., 2006, HEAD Meeting Bullettin, 38, 380
\bibitem[]{} Sato, G., Yamazaki, R., Ioka, K., 2007, ApJ, 657, 359
\bibitem[]{} Schaefer, B. E., 2007, ApJ, 660, 16
\bibitem[]{} Stamatikos, M., Barbier, L., Barthelmy, S. D., 2006, GCN 5639
\bibitem[]{} Tagliaferri, G., Antonelli, L. A., Chincarini, G., et al., 2005, A\&A, 443, L1
\bibitem[]{} Ulanov, M. V., Golenetskii, S. V., Frederiks, D. D., et al.,
  2005, NCimC, 20, 351
\bibitem[]{} Vetere, M.L., et al. 2007, A\&A, 473, 347
\bibitem[]{} Willingale, R., O'Brien, P.T., Osborne, J.P., et al., 2007, ApJ, 662, 1093
\bibitem[]{} Yonetoku, D., Marakami, T., Nakamura, T., et al., 2004, ApJ, 609, 935
\end{thebibliography}
\end{document}